# Mass Acquisition of Dirac Fermions in $Bi_4I_4$ by Spontaneous Symmetry Breaking

Ming Yang[1], Wenxuan Zhao[2], Dan Mu[3], Zhijian Shi[1], Jingyuan Zhong[1], Yaqi Li[1], Yundan Liu[3,*], Jianxin Zhong[4,3], Ningyan Cheng[5], Wei Zhou[6], Jianfeng Wang[1], Yan Shi[7], Ying Sun[1], Weichang Hao[1], Lexian Yang[2], Jincheng Zhuang[1,*], Yi Du[1,*]

[1] School of Physics, Beihang University, Haidian District, Beijing 100191, China
[2] State Key Laboratory of Low Dimensional Quantum Physics, Department of Physics, Tsinghua University, Beijing 100084, China
[3] Hunan Key Laboratory of Micro-Nano Energy Materials and Devices, and School of Physics and Optoelectronics, Xiangtan University, Hunan 411105, China
[4] Institute for Quantum Science and Technology, Shanghai University, Shanghai 200444, China
[5] Information Materials and Intelligent Sensing Laboratory of Anhui Province, Key Laboratory of Structure and Functional Regulation of Hybrid Materials of Ministry of Education, Institutes of Physical Science and Information Technology, Anhui University, Hefei 230601, China
[6] School of Electronic and Information Engineering, Changshu Institute of Technology, Changshu 215500, China
[7] School of Automation Science and Electrical Engineering, Beihang University, Beijing 100191, China

Ming Yang, Wenxuan Zhao, Dan Mu, and Zhijian Shi contributed equally to this work.
*Correspondence authors. E-mail: jincheng@buaa.edu.cn; liuyd@xtu.edu.cn; yi_du@buaa.edu.cn

## Abstract

**Massive Dirac fermions, which are essential for realizing novel topological phenomena, are expected to be generated from massless Dirac fermions by breaking the related symmetry, such as time-reversal symmetry (TRS) in topological insulators or crystal symmetry in topological crystalline insulators. Here, we report scanning tunneling microscopy and angle-resolved photoemission spectroscopy studies of $α$-$Bi_4I_4$, which reveals the realization of massive Dirac fermions in the (100) surface states without breaking the TRS. Combined with first-principle calculations, our experimental results indicate that the spontaneous symmetry breaking engenders two nondegenerate edges states at the opposite sides of monolayer $Bi_4I_4$ after the structural phase transition, imparting mass to the Dirac fermions after taking the interlayer coupling into account. Our results not only demonstrate the formation of the massive Dirac fermions by spontaneous symmetry breaking, but also imply the potential for the engineering of Dirac fermions for device applications.**

The topological insulators (TI) and topologic crystal insulators (TCI) host linearly dispersed boundary bands with massless Dirac fermions and spin texture protected by time-reversal symmetry (TRS) and crystal symmetry (CS), respectively [1–5]. Massive Dirac fermions, which could be valuable for the generation of new topological quantum phenomena and applications of low-consumption spintronics, have been theoretically predicted and experimentally identified in topological surface states by importing perturbations to break TRS in TIs or CS in TCIs [5–12]. Another means to impart mass to electrons is through spontaneous symmetry breaking (SSB), which is akin to the Higgs mechanism in particle physics [9]. One generation process for massive Dirac fermions is shown in the schematic illustration in Fig. 1(a). The massless Dirac fermions reside in the degenerate edge states of a quantum spin Hall insulator (QSHI) with in-plane inversion symmetry (IS). The edges states of the two opposite sides become non-degenerate when the in-plane IS is destroyed by the SSB. When these non-degenerate QSHIs stack along the out-of-plane direction with the inter-plane IS, the massive Dirac fermions are formed under the finite interlayer coupling strength. Nevertheless, the SSB-induced massive Dirac fermions has not been well studied so far, owing to the lack of a suitable material platform.

$Bi_4I_4$, which is constructed from one-dimensional (1D) chains, possesses a room-temperature structural phase transition from $\beta$ phase at high temperature to $\alpha$ phase at low temperature [13-21]. The $\beta$ phase, which possesses in-plane IS and inter-plane IS (inversion centers (IC), labelled by the red stars in Fig. 1(b)), is the first experimentally discovered weak topological insulator with band inversions at two time-reversal-invariant (TRI) momenta [13,22]. The in-plane IS is broken in $\alpha$ phase, as displayed in Fig. 1(c), implying that the SSB takes place in this material. Furthermore, the previous calculations predict the possible QSHI nature of monolayer $Bi_4I_4$, which still needs to be experimentally confirmed [23]. Thus, $\alpha$-$Bi_4I_4$ is an ideal material platform to address the role of SSB in the generation of massive Dirac fermions, as depicted by the schematic diagram in Fig. 1(a).

In this work, we have investigated the electronic structures of $\alpha$-$Bi_4I_4$ using scanning tunneling microscopy/spectroscopy (STM/STS), laser-based angle-resolved photoemission spectroscopy (ARPES), and first-principle calculations. The STS spectra and the calculations reveal the QSHI nature of the monolayer $\alpha$-$Bi_4I_4$ with the non-degenerate edge states in two opposite sides along the chain direction, which is evoked by SSB. The ARPES results calculations demonstrate the formation of the massive Dirac fermions by the mutual effect of SSB and inter-plane interaction.

There is a lattice shift of $b/2$ between every two adjacent layers in $\alpha$ phase compared to $\beta$ phase, leading to the double unit cell volume in the $\alpha$ phase and the correlated structural phase transition, as shown in Fig. 1(b) and (c). The details on the phase transition and structural parameters can be obtained in [24] Compared to the $\beta$ phase at high temperature, $\alpha$ phase at low temperature has less ICs, which are labelled by red stars in Fig. 1(b) and (c) due to the lattice distortion after a structural transition, which is a typical SSB in this system. The structural characterizations by high-angle annular dark-field scanning transmission electron microscope (HAADF-STEM) in Fig. 1(d) and X-ray diffraction

(XRD) spectrum in Fig. 1(e) indicate the high-quality of our sample.

Our first-principles calculations show band inversion with reversal of parities at the $M$ point of the Brillouin zone (BZ) for both the monolayer $\alpha$ phase and the $\beta$ phase, as displayed in Fig. 2(a) and [24], indicating the QSHI characteristic of the two monolayer phases with a small difference in gap value due to the structural discrepancy. STM, which is a powerful tool used to elucidate the bulk-boundary correspondence on the nanoscale, has been applied to demonstrate the topological non-trivial edge states in various QSHIs [11,30–43]. A large-scale STM image and a high-resolution STM image of the cleaved (001) plane of $\alpha$-$Bi_4I_4$ are displayed in Fig. 2(b) and (c), respectively, which is similar to the STM results for $Bi_4Br_4$ in previous reports [42,44]. The STS curves collected at different distances across the edge are displayed in Fig. 2(d), where the semiconducting nature of the two-dimensional (2D) upper terrace and lower terrace is revealed by the energy gap. There is a difference in the gap value between two adjacent layers, which is caused by the discrepancy in the relaxed structures between two different exposed surfaces [24]. Moreover, the STS curves collected at edge positions show the nearly constant density of states (DOS) residing in the gap region, coinciding with the linear Dirac cone dispersion in 1D. Our STS results coincide with the calculations in Fig. S5 and demonstrate the QSHI nature of monolayer $\alpha$-$Bi_4I_4$.

We applied density functional theory (DFT) calculations on the band structure of three-dimensional (3D) $\alpha$-$Bi_4I_4$ and laser-based ARPES measurements on the (100) plane to resolve this debate on the topological nature of bulk $\alpha$-$Bi_4I_4$. The first-principles calculations show the band inversions take place twice at both the $M$ and $L$ points of the 3D BZ. The parity eigenvalues below $E_F$, which are linked to symmetry-indicator topological invariants [2,45], are retained after the double band inversion at two TRI momenta, resulting in the trivial topological invariants [24]. Fig. 3(b) displays the constant energy contours (CECs) of the (100) surface measured by laser-ARPES at 77 K with different binding energies. The linear-like band dispersion could be identified in the CEC extending along the $\bar{\Gamma} - \bar{M}$ direction, which is ascribed to the quasi-1D lattice structure of this system [13–23,41–44,46–56]. Fig. 3(c) and (f) shows the dispersions along the chain direction at the two TRI momenta of the (100) surface BZ, at $\bar{\Gamma}$ and $\bar{Z}$, corresponding to cut 1 and cut 2 labelled in Fig. 3(b), respectively. The bands away from $E_F$ with high spectral weight and the linear dispersion near $E_F$ are identified as the bulk valence bands (BVB) from two adjacent layers and surface states (SS1 and SS2), respectively [14]. Interestingly, we can observe the splitting of the surface states from the enlarged view in Fig. 3(d), (e), (g), and (i), which nicely agrees with the massive Dirac fermions, as depicted in Fig. 1(a), but in drastic contrast to the single Dirac cone-like band in Ref. [14] or the absence of surface states in Ref. [13]. Moreover, the second derivatives of Fig. 3(d) and (h) are displayed in Fig. 3(f) and (i), respectively, where the two nondegenerate surface states arising from two intersected directional Dirac cones with the hybridized energy gap can be clearly identified. Our ARPES spectra show compelling evidence for the formation of quasi-1D massive Dirac fermions in the surface states of the (100) plane of $\alpha$-$Bi_4I_4$.

In order to determine the formation mechanism of the (100) surface states with massive Dirac fermions, we performed band structure calculations on monolayer $\beta$-Bi$_4$I$_4$, monolayer $\alpha$-Bi$_4$I$_4$, 3D $\beta$-Bi$_4$I$_4$, and 3D $\alpha$-Bi$_4$I$_4$ based on Green function methods. For the monolayer $\beta$-Bi$_4$I$_4$, the calculated results show the degenerate edge states of the two opposite sides, as shown in Fig. 4(b), protected by the in-plane IS marked by the red star in Fig. 4(a). This in-plane IS is broken in monolayer $\alpha$-Bi$_4$I$_4$ (Fig. 4(c)), *i.e.* SSB, giving rise to the non-degenerate edge states of the two opposite sides (Fig. 4(d)). The non-degenerate nature of the edge states has been confirmed by a comparative study between the calculated DOS of the two opposite sides and the differential conductance (d$I$/d$V$) curves of edge states collected at different edges, as displayed in Fig. 2(e), where the same DOS divergence of the edge states of two adjacent layers can be clearly explored. It is noticeable that all the previous work on the investigation of QSHI reveal the surrounding degenerate topological edge states, and our work exhibits a novel QSHI with two non-degenerate edge states of the opposite sides for the first time in experiment. The 3D $\beta$-Bi$_4$I$_4$ is constructed by the stacking of the monolayer $\beta$-Bi$_4$I$_4$ with both in-plane IS and inter-plane IS (See Fig. 4(e)), where the degenerate 1D edge states are also arranged along the lattice $c$ axis to form the 2D surface states with finite interlayer coupling strength. The calculated energy-momentum ($E$-$k$) dispersions cut along the TRI momenta are displayed in Fig. 4(f) along with the experimental results, where the gapless Dirac cone correlated to a weak topological insulator can be clearly observed. It should be noted that there is an energy shift of the Dirac points of these two $E$-$k$ dispersions, which is a reflection of the interlayer interaction. The role of interlayer interaction strength can be founded in [24]. The 3D $\alpha$-Bi$_4$I$_4$ is built from stacking monolayer $\alpha$-Bi$_4$I$_4$ with only inter-plane IS. This inter-plane IS offers the same edge states of the left (right) side edge of the B layer and the right (left) side edge of the A layer, as shown in Fig. 4(g). Consequently, these two edge states are alternatively arranged in the (100) surface to form the 2D surface states. Our calculations in Fig. 4(h) imply that the two non-degenerate Dirac cones derived from the two non-degenerate edge states in Fig. 4(d) intersect each other at the points deviating from TRI momenta, which is well consistent with the laser-ARPES results, as shown in Fig. 4(h). Therefore, the SS1 and SS2 observed in Fig. 3 are derived from the two non-degenerate Dirac cones of topological edge states. Unlike 3D $\beta$-Bi$_4$I$_4$, the role of interlayer interaction in the band structure of 3D $\alpha$-Bi$_4$I$_4$ is expressed by the opening of the energy gaps at the crossing points, imparting mass to the Dirac fermions. Our simulation results well reproduce the experimental features measured by ARPES and reveal the formation mechanism of massive Dirac fermions in this system. Moreover, the interlayer coupling can also evoke the energy gap in the edge state of double-layer step edge similar to the (100) surface states of $\alpha$-Bi$_4$I$_4$ but in contrast to gapless feature in the monolayer case, provides a flexible material platform to realize the switch of energy gap in the Dirac bands by varying the layer numbers.

In conclusion, we reveal a novel QSHI with two non-degenerate edge states on the opposite sides in monolayer $\alpha$-Bi$_4$I$_4$, which is different from the previous QSHI with surrounding degenerate edge states. These two non-degenerate edge states are alternatively arranged along the lattice $c$-direction to form the (100) surface states. The thus-formed surface states host massive Dirac fermions in the (100) plane of a topological trivial insulator (3D $\alpha$-Bi$_4$I$_4$), for which the origin is the SSB with the support of interlayer coupling. Our work provides compelling evidence that spin-polarized gapped surface states can be formed regardless of the topological nature of the material as determined by the topological band theory [24].

**Acknowledgements**

This work was supported by the National Natural Science Foundation of China (12274016, 12074021, 12004321, 11904015, and 52473287), the National Key R&D Program of China (2018YFE0202700, 2022YFB3403400, and 2022YFB3403401), and the Fundamental Research Funds for the Central Universities (Grant No. YWF-23SD00-001 and YWF-22-K-101).

# Reference

1. B. A. Bernevig, T. L. Hughes, and S.-C. Zhang, Quantum Spin Hall Effect and Topological Phase Transition in HgTe Quantum Wells. Science **314**, 1757 (2006).
2. L. Fu and C. L. Kane, Topological Insulators with Inversion Symmetry. Phys. Rev. B **76**, 045302 (2007).
3. L. Fu, Topological Crystalline Insulators. Phys. Rev. Lett. **106**, 106802 (2011).
4. T. H. Hsieh, H. Lin, J. Liu, W. Duan, A. Bansil, and L. Fu, Topological Crystalline Insulators in the SnTe Material Class. Nat. Commun. **3**, 1 (2012).
5. Y. Okada et al., Observation of Dirac Node Formation and Mass Acquisition in a Topological Crystalline Insulator, Science **341**, 1496 (2013).
6. X.-L. Qi, T. L. Hughes, and S.-C. Zhang, Topological Field Theory of Time-Reversal Invariant Insulators, Phys. Rev. B **78**, 195424 (2008).
7. X.-L. Qi, R. Li, J. Zang, and S.-C. Zhang, Inducing a Magnetic Monopole with Topological Surface States, Science **323**, 1184 (2009).
8. Y. L. Chen, et al., Massive Dirac Fermion on the Surface of a Magnetically Doped Topological Insulator, Science **329**, 659 (2010).
9. T. Sato, K. Segawa, K. Kosaka, S. Souma, K. Nakayama, K. Eto, T. Minami, Y. Ando, and T. Takahashi, Unexpected Mass Acquisition of Dirac Fermions at the Quantum Phase Transition of a Topological Insulator, Nat. Phys. **7**, 11 (2011).
10. M. Serbyn and L. Fu, Symmetry Breaking and Landau Quantization in Topological Crystalline Insulators, Phys. Rev. B **90**, 035402 (2014).
11. I. Zeljkovic et al., Dirac Mass Generation from Crystal Symmetry Breaking on the Surfaces of Topological Crystalline Insulators, Nat. Mater. **14**, 3 (2015).
12. L. Ye et al., Massive Dirac Fermions in a Ferromagnetic Kagome Metal, Nature **555**, 7698 (2018).
13. R. Noguchi et al., A Weak Topological Insulator State in Quasi-One-Dimensional Bismuth Iodide, Nature **566**, 518 (2019).
14. J. Huang et al., Room-Temperature Topological Phase Transition in Quasi-One-Dimensional Material $Bi_4I_4$, Phys. Rev. X **11**, 031042 (2021).
15. G. Autès et al., A Novel Quasi-One-Dimensional Topological Insulator in Bismuth Iodide β-$Bi_4I_4$, Nature Mater **15**, 2 (2016).
16. A. Pisoni, R. Gaál, A. Zeugner, V. Falkowski, A. Isaeva, H. Huppertz, G. Autès, O. V. Yazyev, and L. Forró, Pressure Effect and Superconductivity in the β−$Bi_4I_4$ Topological Insulator, Phys. Rev. B **95**, 235149 (2017).
17. Y. Qi et al. Pressure-induced Superconductivity and Topological Quantum Phase Transitions in a Quasi-One-Dimensional Topological Insulator: $Bi_4I_4$. npj Quant. Mater. **3**, 4 (2018).
18. X. Wang, J. Wu, J. Wang, T. Chen, H. Gao, P. Lu, Q. Chen, C. Ding, J. Wen, and J. Sun, Pressure-Induced Structural and Electronic Transitions in Bismuth Iodide, Phys. Rev. B **98**, 174112 (2018).
19. D.-Y. Chen et al., Quantum transport properties in single crystals of α-$Bi_4I_4$. Phys. Rev. Mater. **2**, 114408 (2018).
20. P. Wang, F. Tang, P. Wang, H. Zhu, C.-W. Cho, J. Wang, X. Du, Y. Shao, and L. Zhang, Quantum Transport Properties of β-$Bi_4I_4$ Near and Well beyond the Extreme Quantum Limit, Phys. Rev. B **103**, 155201 (2021).

21. Y. Liu et al., Gate-Tunable Transport in Quasi-One-Dimensional α-Bi$_4$I$_4$ Field Effect Transistors, Nano Lett. **22**, 1151 (2022).
22. C.-C. Liu, J.-J. Zhou, Y. Yao, and F. Zhang, Weak Topological Insulators and Composite Weyl Semimetals: β-Bi$_4$X$_4$ (X = Br, I). Phys. Rev. Lett. **116**, 066801 (2016).
23. J.-J. Zhou, W. Feng, C.-C. Liu, S. Guan, and Y. Yao, Large-Gap Quantum Spin Hall Insulator in Single Layer Bismuth Monobromide Bi$_4$Br$_4$, Nano Lett. **14**, 4767 (2014).
24. See Supplemental Material, which includes Refs. [25-29], for additional information about the methods, experimental electronic properties, and calculated band structures.
25. G. Kresse and J. Furthmüller, Efficient Iterative Schemes for Ab Initio Total-Energy Calculations Using a Plane-Wave Basis Set, Phys. Rev. B **54**, 11169 (1996).
26. P. E. Blöchl, Projector Augmented-Wave Method, Phys. Rev. B **50**, 17953 (1994).
27. J. P. Perdew, K. Burke, and M. Ernzerhof, Generalized Gradient Approximation Made Simple, Phys. Rev. Lett. **77**, 3865 (1996).
28. I. Souza, N. Marzari, and D. Vanderbilt, Maximally Localized Wannier Functions for Entangled Energy Bands, Phys. Rev. B **65**, 035109 (2001).
29. Q. Wu, S. Zhang, H.-F. Song, M. Troyer, and A. A. Soluyanov, WannierTools: An Open-Source Software Package for Novel Topological Materials, Comput. Phys. Commun. **224**, 405 (2018).
30. C. Pauly, B. Rasche, K. Koepernik, M. Liebmann, M. Pratzer, M. Richter, J. Kellner, M. Eschbach, B. Kaufmann, and L. Plucinski, Subnanometre-Wide Electron Channels Protected by Topology, Nat. Phys. **11**, 338–343 (2015).
31. X.-B. Li et al., Experimental Observation of Topological Edge States at the Surface Step Edge of the Topological Insulator ZrTe$_5$. Phys. Rev. Lett. **116**, 176803 (2016).
32. Wang, Z. F. et al. Topological edge states in a high-temperature superconductor FeSe/SrTiO$_3$(001) film. *Nat. Mater.* **15**, 968–973 (2016).
33. R. Wu, J.-Z. Ma, S.-M. Nie, L.-X. Zhao, X. Huang, J.-X. Yin, B.-B. Fu, P. Richard, G.-F. Chen, and Z. Fang, Evidence for Topological Edge States in a Large Energy Gap near the Step Edges on the Surface of ZrTe$_5$. Phys. Rev. X **6**, 021017 (2016).
34. F. Reis, G. Li, L. Dudy, M. Bauernfeind, S. Glass, W. Hanke, R. Thomale, J. Schäfer, and R. Claessen, Bismuthene on a SiC Substrate: A Candidate for a High-Temperature Quantum Spin Hall Material, Science **357**, 287 (2017).
35. L. Peng, Y. Yuan, G. Li, X. Yang, J.-J. Xian, C.-J. Yi, Y.-G. Shi, and Y.-S. Fu, Observation of Topological States Residing at Step Edges of WTe$_2$, Nat. Commun. **8**, 659 (2017).
36. S. Tang et al., Quantum Spin Hall State in Monolayer 1T′-WTe$_2$. Nat. Phys. **13**, 683–687 (2017).
37. M. M. Ugeda, A. Pulkin, S. Tang, H. Ryu, Q. Wu, Y. Zhang, D. Wong, Z. Pedramrazi, A. Martín-Recio, and Y. Chen, Observation of Topologically Protected States at Crystalline Phase Boundaries in Single-Layer WSe$_2$, Nat. Commun. **9**, 3401 (2018).
38. X. Dong et al., Observation of Topological Edge States at the Step Edges on the Surface of Type-II Weyl Semimetal TaIrTe$_4$. ACS Nano **13**, 9571 (2019).
39. R. Stühler, F. Reis, T. Müller, T. Helbig, T. Schwemmer, R. Thomale, J. Schäfer, and R. Claessen, Tomonaga–Luttinger Liquid in the Edge Channels of a Quantum Spin Hall Insulator, Nat. Phys. **16**, 1 (2020).
40. N. Avraham et al., Visualizing Coexisting Surface States in the Weak and Crystalline Topological Insulator Bi$_2$TeI. Nat. Mater. **19**, 610 (2020).
41. J. Zhuang, J. Li, Y. Liu, D. Mu, M. Yang, Y. Liu, W. Zhou, W. Hao, J. Zhong, and Y. Du, Epitaxial Growth of Quasi-One-Dimensional Bismuth-Halide Chains with Atomically Sharp Topological Non-Trivial Edge States. ACS Nano **15**, 14850 (2021).
42. M. Yang et al., Large-Gap Quantum Spin Hall State and Temperature-Induced Lifshitz Transition in Bi$_4$Br$_4$, ACS Nano **16**, 3036 (2022).
43. J. Zhong et al., Towards Layer-selective Quantum Spin Hall Channels in Weak Topological Insulator Bi$_4$Br$_2$I$_2$. Nat. Commun. **14**, 4964 (2023).
44. N. Shumiya et al., Evidence of a Room-Temperature Quantum Spin Hall Edge State in a Higher-Order

Topological Insulator, Nat. Mater. **21**, 10 (2022).

45. E. Khalaf, H. C. Po, A. Vishwanath, and H. Watanabe, Symmetry Indicators and Anomalous Surface States of Topological Crystalline Insulators, Phys. Rev. X **8**, 031070 (2018).
46. R. Noguchi et al., Evidence for a Higher-Order Topological Insulator in a Three-Dimensional Material Built from van der Waals Stacking of Bismuth-Halide Chains, Nat. Mater. **20**, 473 (2021).
47. J. Zhong, M. Yang, F. Ye, C. Liu, J. Wang, J. Wang, W. Hao, J. Zhuang, and Y. Du, Facet-Dependent Electronic Quantum Diffusion in the High-Order Topological Insulator $Bi_4Br_4$, Phys. Rev. Applied **17**, 064017 (2022).
48. W. Zhao et al., Topological Electronic Structure and Spin Texture of Quasi-One-Dimensional Higher-Order Topological Insulator $Bi_4Br_4$. Nat. Commun. **14**, 8089 (2023).
49. J. S. Oh et al., Ideal Weak Topological Insulator and Protected Helical Saddle Points, Phys. Rev. B **108**, L201104 (2023).
50. X. Zhang et al., Controllable Epitaxy of Quasi-One-Dimensional Topological Insulator $α-Bi_4Br_4$ for the Application of Saturable Absorber, Appl. Phys. Lett. **120**, 093103 (2022).
51. D.-Y. Chen, D. Ma, J. Duan, D. Chen, H. Liu, J. Han, and Y. Yao, Quantum Transport Evidence of the Boundary States and Lifshitz Transition in $Bi_4Br_4$. Phys. Rev. B **106**, 075206 (2022).
52. X. Peng et al., Observation of Topological Edge States on $α-Bi_4Br_4$ Nanowires Grown on $TiSe_2$ Substrates. J. Phys. Chem. Lett. **12**, 10465 (2021).
53. X. Li et al., Pressure-Induced Phase Transitions and Superconductivity in a Quasi–1-Dimensional Topological Crystalline Insulator $α-Bi_4Br_4$. Proc. Natl. Acad. Sci. USA **116**, 17696 (2019).
54. C.-H. Hsu, X. Zhou, Q. Ma, N. Gedik, A. Bansil, V. M. Pereira, H. Lin, L. Fu, S.-Y. Xu, and T.-R. Chang, Purely Rotational Symmetry-Protected Topological Crystalline Insulator $α-Bi_4Br_4$. 2D Mater. **6**, 031004 (2019).
55. J.-J. Zhou, W. Feng, G.-B. Liu, and Y. Yao, Topological Edge States in Single- and Multi-Layer. New J. Phys. **17**, 015004 (2015).
56. S. Deng, X. Song, X. Shao, Q. Li, Y. Xie, C. Chen, and Y. Ma, First-Principles Study of High-Pressure Phase Stability and Superconductivity of $Bi_4I_4$, Phys. Rev. B **100**, 224108 (2019).

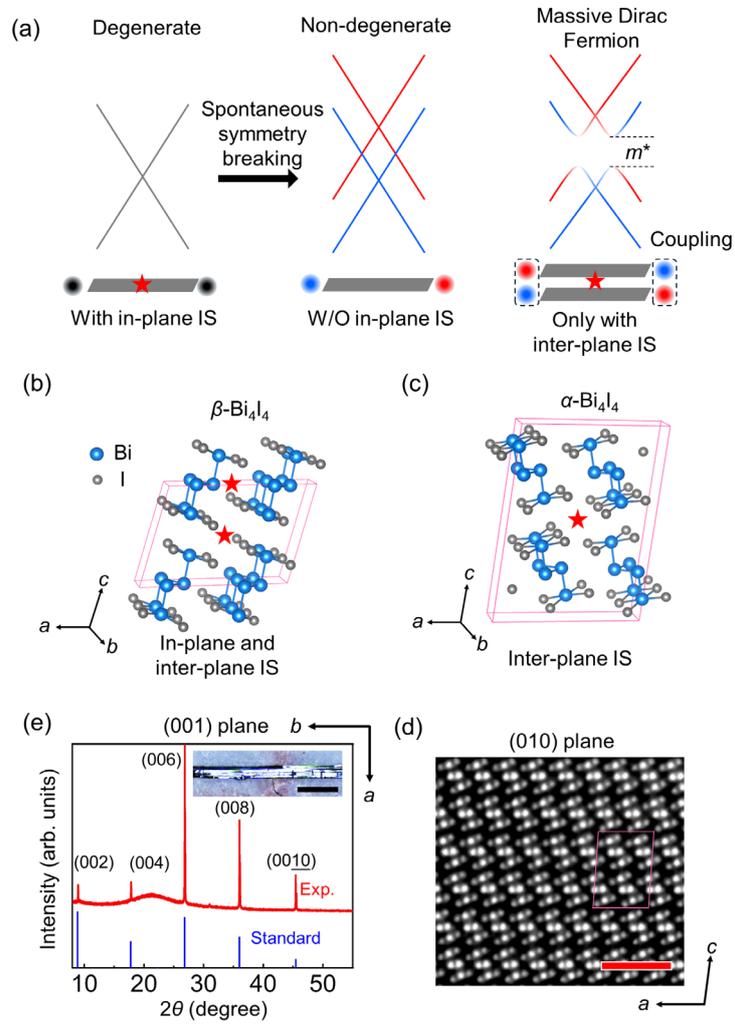

FIG. 1. Crystal and band structure of $Bi_4I_4$. (a), Schematic diagrams of the formation of the massive Dirac fermions by the mutual effect of spontaneous symmetry breaking and interlayer coupling. IS: inversion symmetry. The black solid circles mark the degenerate edge states of QSHI. The red and blue solid circles mark the two non-degenerate edge states. Structural models of (b), $β$-$Bi_4I_4$ with a single-layer unit cell, and (c), $α$-$Bi_4I_4$ with a double-layer unit cell. The blue and grey balls represent Bi (including internal locations ($Bi_{in}$) and external locations ($Bi_{ex}$)) and I atoms, respectively. (d) HAADF-STEM image of (010) plane in $α$-$Bi_4I_4$ with the double-layer unit cell identified by the pink rhomboid. Scale bar, 2 nm. (e), (001) XRD spectrum of $α$-$Bi_4I_4$. Inset: optical image of the (001) surface. Scale bar, 0.5 mm.

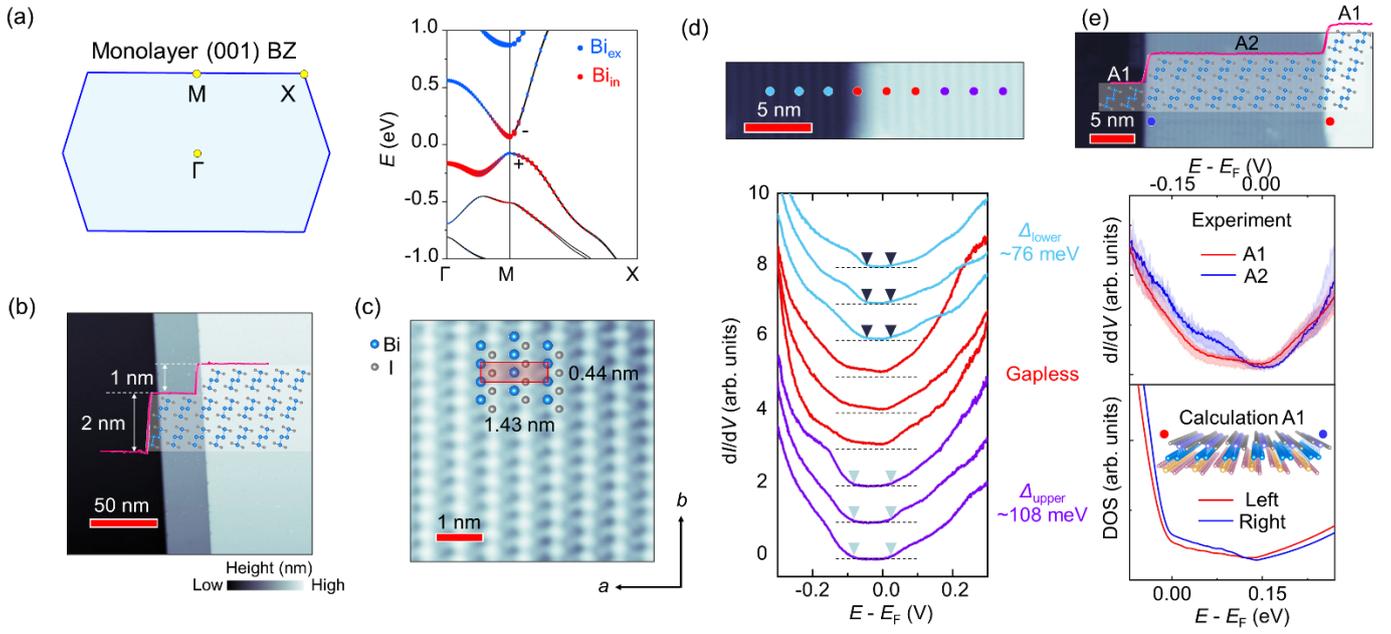

FIG. 2. QSHI nature of monolayer α-Bi₄I₄. (a), Calculated band structure of monolayer α-Bi₄I₄ with orbital projected character, along with the BZ of the primitive cell of monolayer α-Bi₄I₄. The components of $Bi_{ex}$ and $Bi_{in}$ are marked by blue and red dots, respectively, with the correlated parities. (b), Large-scale STM image of the cleaved (001) surface of Bi₄I₄ with the projected step height (red line) and structural model with different steps ($V_{bias}$ = 1 V, $I$ = 100 pA). The side view of crystal structure has been projected to illuminate the arrangement of steps. (c), High-resolution STM image of the (001) surface ($V_{bias}$ = 1 V, $I$ = 1 nA). The red rectangle marks the lattice unit cell of monolayer α-Bi₄I₄. (d), Upper panel: topography of the (001) surface. Lower panel: STS curves collected at the positions marked in the upper panel. The black triangles and blue triangles provide guidance to the eye of the energy gap correlated with the lower terrace and upper terrace, respectively. (e), The top panel shows the double-step (001) surface topography with two adjacent monolayer steps and the projected structural model. The middle and lower panels display a comparison between the experimental and calculated DOS of edge states, which exhibit high consistency with each other. STS results were obtained at the edges of the A1 and A2 layers, as indicated by the red and blue circles, respectively. The light curves are the differential spectra collected at different positions for edge states, and the dark curves denote the related average spectra.

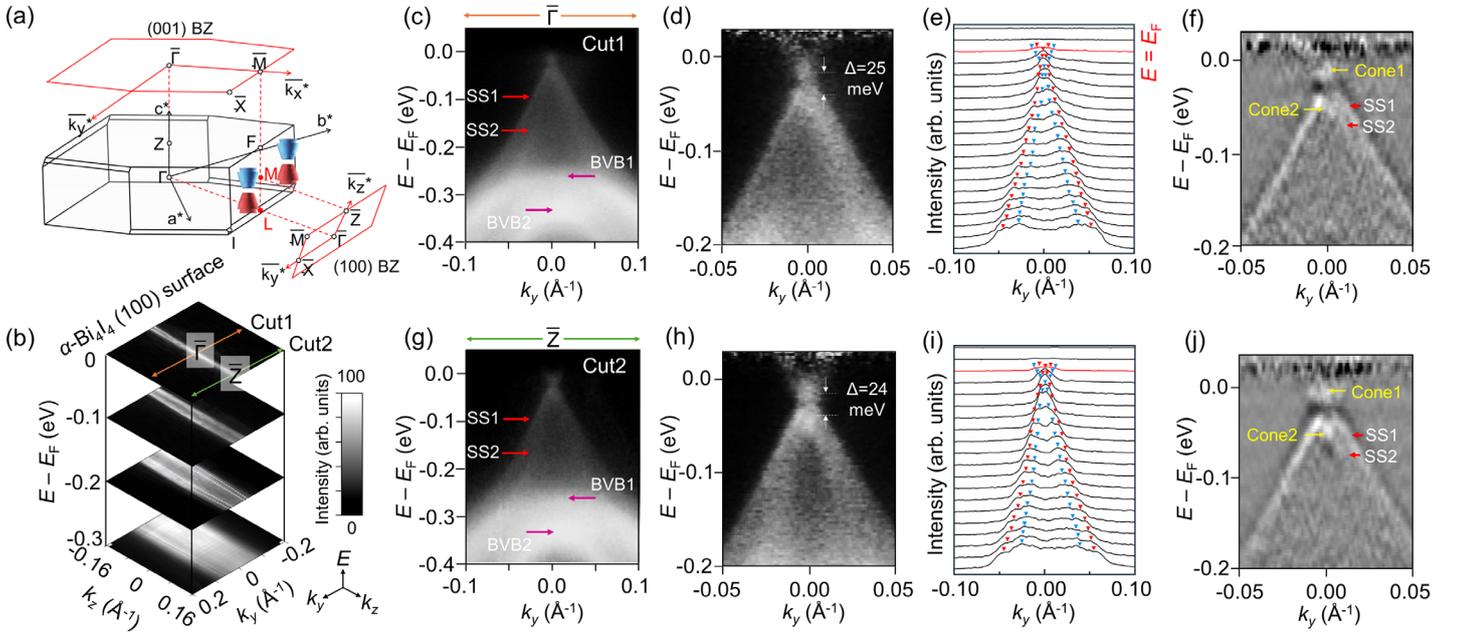

FIG. 3. Band structure of 3D α-Bi$_4$I$_4$. (a), 3D BZ with (001) and (100) projected BZs. The schematic diagram shows the locations of band inversion (*M* point and *L* point). (b), CECs of the α-Bi$_4$I$_4$ (100) surface at different binding energies. Two pairs of quasi-linear bands are denoted by white dots. Large-energy-range *E*-*k$_y$* dispersion spectra at (c), $\bar{\Gamma}$ and (g), $\bar{Z}$, respectively. Double pairs of BVBs and surface states (SSs) are indicated by pink and red arrows, respectively. Small-energy-range *E*-*k$_y$* dispersion spectra at (d), $\bar{\Gamma}$ and (h), $\bar{Z}$, respectively. Momentum distribution curves (MDCs) with two pairs of SSs denoted by blue and red triangles at $\bar{\Gamma}$ and $\bar{Z}$ are shown at (e) and (i), respectively. The coupling gaps can be resolved as 25 meV and 24 meV. Quadratic differential spectra at (f), $\bar{\Gamma}$ and (j), $\bar{Z}$, respectively. Two Dirac cones originating from edge states of the QSHI are denoted as Cone1 (higher energy) and Cone2 (lower energy).

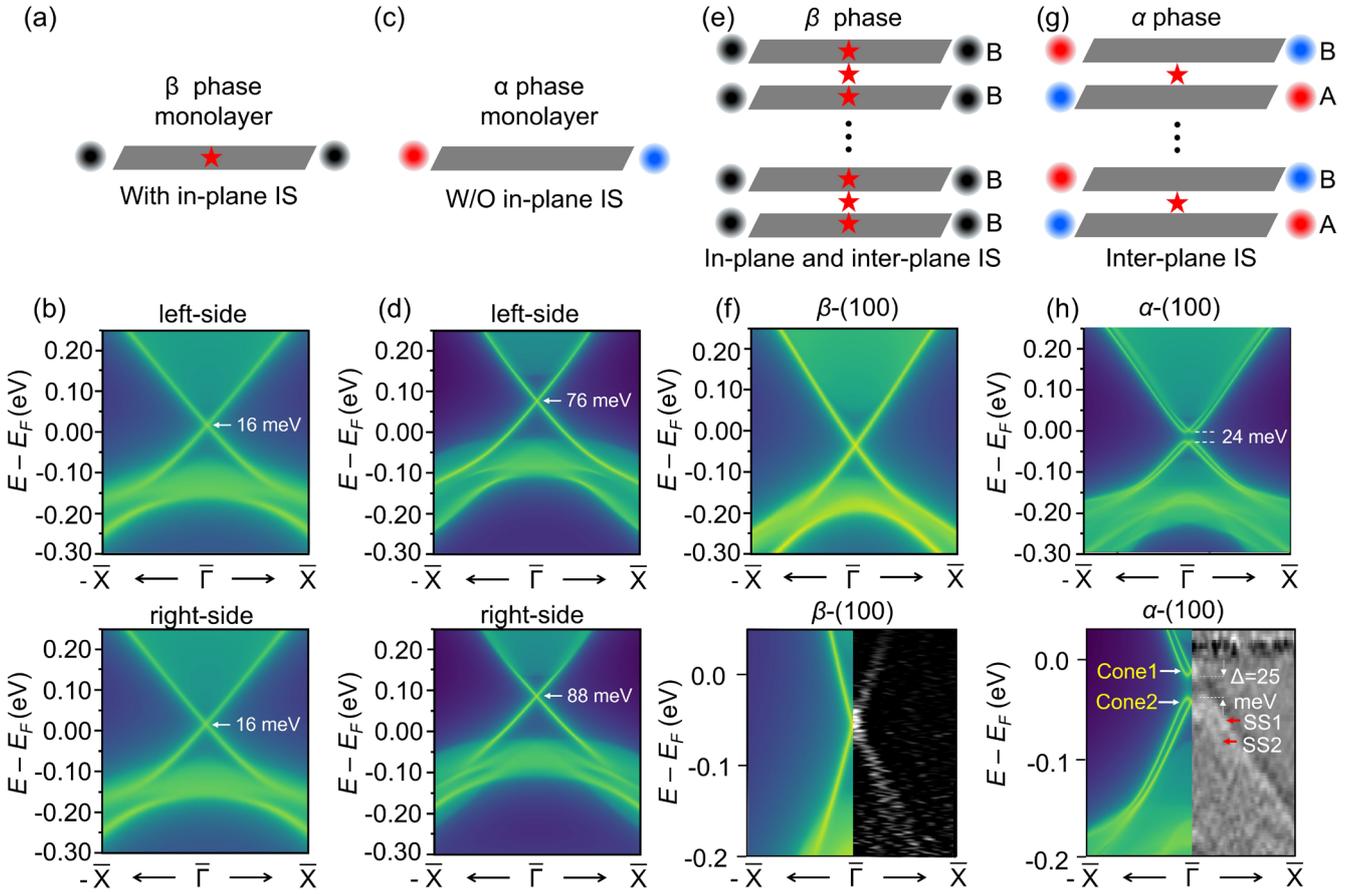

FIG. 4. Formation of massive Dirac fermions. Schematic evolution of the edge states of (a), monolayer $\beta$-Bi$_4$I$_4$, (c), monolayer $\alpha$-Bi$_4$I$_4$, (e), 3D $\beta$-Bi$_4$I$_4$, and (g), 3D $\alpha$-Bi$_4$I$_4$ with the IS labelled by red stars. The black solid circles mark the degenerate edge states of the QSHI. The red and blue solid circles denote the two non-degenerate edge states. Calculated left-hand-side boundary and right-hand-side boundary of (b), monolayer $\beta$-Bi$_4$I$_4$ and (d), monolayer $\alpha$-Bi$_4$I$_4$, and the (100) surface state of 3D (f), $\beta$-Bi$_4$I$_4$ and (h), $\alpha$-Bi$_4$I$_4$ along the $\overline{\Gamma}$-$\overline{X}$ direction in the projected surface BZ. The lower panels of (f) and (h) are the comparisons of the calculations and the ARPES spectra of the (100) surface of $\beta$-Bi$_4$I$_4$ and $\alpha$-Bi$_4$I$_4$, respectively.